\documentclass[useAMS,usenatbib,twocolumn]{mn2e}
\usepackage{amsmath}
\usepackage{dcolumn}
\usepackage[utf8]{inputenc}
\usepackage{graphics,graphicx}
\usepackage{float}

\usepackage{color}

\newcommand{\p}{^\prime}
\newcommand{\pp}{^{\prime\prime}}

\title[Hydronium probes of a variable $\mu$]{Accurate prediction of H$_3$O$^+$ and D$_3$O$^+$ sensitivity coefficients to probe a variable proton-to-electron mass ratio}

\date{\today}

\author[A. Owens et al.]
{A. Owens$^{1,2}$, S. N. Yurchenko$^2$, O. L. Polyansky$^2$, R. I. Ovsyannikov$^3$, W. Thiel$^1$ \newauthor and V. \v{S}pirko$^{4,5}$\thanks{The corresponding author: \texttt{spirko@marge.uochb.cas.cz}} \\ \\
$^1$ Max-Planck-Institut f\"{u}r Kohlenforschung, Kaiser-Wilhelm-Platz 1, 45470 M\"{u}lheim an der Ruhr, Germany\\
$^2$ Department of Physics and Astronomy, University College London, Gower Street, WC1E 6BT London, United Kingdom\\
$^3$ Institute of Applied Physics, Russian Academy of Sciences, Nizhny Novgorod, 603950 Russia \\
$^4$ Academy of Sciences of the Czech Republic, Institute of Organic Chemistry and Biochemistry,\\
Flemingovo n\'am.~2, 166 10 Prague 6, Czech Republic \\
$^5$ Department of Chemical Physics and Optics, Faculty of Mathematics and Physics, Charles University in Prague,\\
Ke Karlovu 3, CZ-12116 Prague 2, Czech Republic}

\date{Accepted XXXX. Received XXXX; in original form XXXX}

\pagerange{\pageref{firstpage}--\pageref{lastpage}} \pubyear{2015}

\begin{document}

\label{firstpage}

\maketitle

\begin{abstract}
The mass sensitivity of the vibration-rotation-inversion transitions of H$_3{}^{16}$O$^+$, H$_3{}^{18}$O$^+$, and D$_3{}^{16}$O$^+$ is investigated variationally using the nuclear motion program TROVE~\citep{TROVE:2007}. The calculations utilize new high-level \textit{ab initio} potential energy and dipole moment surfaces. Along with the mass dependence, frequency data and Einstein A coefficients are computed for all transitions probed. Particular attention is paid to the $\Delta|k|=3$ and $\Delta|k-l|=3$ transitions comprising the accidentally coinciding $|J,K\!=\!0,v_2\!=\!0^+\rangle$ and $|J,K\!=\!3,v_2\!=\!0^-\rangle$ rotation-inversion energy levels. The newly computed probes exhibit sensitivities comparable to their ammonia and methanol counterparts, thus demonstrating their potential for testing the cosmological stability of the proton-to-electron mass ratio. The theoretical TROVE results are in close agreement with sensitivities obtained using the nonrigid and rigid inverter approximate models, confirming that the \textit{ab initio} theory used in the present study is adequate.
\end{abstract}


\section{Introduction}

   The hydronium cation (H$_3$O$^+$) is one of the key molecular ions for inferring properties of the interstellar medium, particularly for constraining the cosmic-ray ionization rate of atomic and molecular hydrogen (see \citet{Indriolo:2015} and references therein). Knowledge of such parameters is of astrophysical importance, and as a result, H$_3$O$^+$ is one of the most searched for galactic and extragalactic interstellar molecules~\citep{Hollis:1986,Wootten:1986,Wootten:1991,Phillips:1992,Boreiko:1993,Goi:2001,Tak:2006,Tak:2008,Gerin:2010,Gupta:2010,Aalto:2011,Gonzalez:2013,Lis:2014}. Since H$_3$O$^+$ formation requires presence of H$_2$O, and the chemical relation between H$_3$O$^+$ and H$_2$O is well-understood, H$_3$O$^+$ can serve as an excellent proxy for H$_2$O, which is often hard to observe directly~\citep{Timmermann:1996}. 
   
   Similar to the ammonia molecule, H$_3$O$^+$ has several far infrared (FIR) and submillimetre transitions that are particularly sensitive to the proton-to-electron mass ratio $\mu$~\citep{Kozlov:2011a,Kozlov:2011b}. The most robust constraint on a variable $\mu$ has recently been determined using methanol absorption spectra observed in the lensing galaxy PKS1830$-$211~\citep{Kanekar:2015}. The three measured lines possessed sensitivities differing by $\Delta T=6.4$, where $T$ is the sensitivity coefficient of a transition. In principle then, hydronium is capable of being used exclusively to constrain a possible variation in the proton-to-electron mass ratio, thus avoiding certain systematic errors which arise when using transitions from different molecular species~\citep{Flambaum:2007,Murphy:2008,Henkel:2009,Kanekar:2011}.
   
   A small number of pure inversion and rotation-inversion transitions in the ground vibrational state of H$_3$O$^+$ were originally investigated by \citet{Kozlov:2011a}. However the calculated sensitivity coefficients were overestimated and new values have been computed for H$_3$O$^+$, along with the isotopologues H$_2$DO$^+$, HD$_2$O$^+$, and D$_3$O$^+$~\citep{Kozlov:2011b}. Given the astronomical relevance of H$_3$O$^+$, and a good representative set of accurately measured experimental data~\citep{Uy:1997,Tang:1999,Araki:1999,Furuya:2005,Yu:2009,Yu:2014}, we find it worthwhile to carry out a comprehensive study of hydronium, H$_3{}^{16}$O$^+$ (also referred to as H$_3$O$^+$), and its two symmetric top isotopologues, H$_3{}^{18}$O$^+$ and D$_3{}^{16}$O$^+$. To do this we employ a highly accurate variational approach, which was recently applied to ammonia~\citep{Owens:2015}. Like NH$_3$~\citep{Jansen:2014,Spirko:2014,Owens:2015}, there is a possibility to find transitions with strongly anomalous sensitivities caused by the $\Delta k=\pm 3$ interactions (see \citet{Papousek:1986}), which have not yet been considered.
   
\section{Variational Approach}

   To calculate sensitivity coefficients we follow the same approach that was employed for ammonia~\citep{Owens:2015}. The key assumption is that all baryonic matter may be treated equally~\citep{Dent:2007}, and so $\mu$ is assumed to be proportional to the molecular mass. One can then use suitably scaled values for the mass of hydronium and perform a series of calculations, from which numerical values of the required derivatives, ${\rm d}E/{\rm d}\mu$, can be obtained. The sensitivity coefficient $T_{u,l}$ is defined as
\begin{equation}
T_{u,l}=\frac{\mu}{E_u-E_l}\Bigl(\frac{{\rm d}E_u}{{\rm d}\mu}-\frac{{\rm d}E_l}{{\rm d}\mu}\Bigr),
\label{eq.T}
\end{equation}
where $E_u$ and $E_l$ refer to the energy of the upper and lower state, respectively. The resulting sensitivities can then be used to determine the induced frequency shift of a probed transition, given by the expression
\begin{equation}
\frac{\Delta\nu}{\nu_0}=T_{u,l}\frac{\Delta\mu}{\mu_0},
\label{eq.shift}
\end{equation}
where $\Delta\nu=\nu_{\mathrm{obs}}-\nu_0$ is the change in the frequency, and $\Delta\mu=\mu_{\mathrm{obs}}-\mu_0$ is the change in $\mu$, both with respect to their present day values $\nu_0$ and $\mu_0$.

   Calculations were carried out using the nuclear motion code TROVE~\citep{TROVE:2007}. To compute ro-vibrational transitions and corresponding intensities~\citep{09YuBaYa.NH3}, TROVE requires as input a potential energy surface (PES) and dipole moment surface (DMS). For the present study, new high-level \textit{ab initio} PES and DMS have been utilized. A detailed description of these will be reported elsewhere~(Polyansky \& Ovsyannikov (in preparation)). Here we only provide a summary of the \textit{ab initio} calculations used to generate the respective surfaces. 
   
   The PES was computed at the all-electron multireference configuration interaction (MRCI) level of theory using the core-valence-weighted basis sets, aug-cc-pwCVQZ and aug-cc-pwCV5Z. A two-point formula was applied to extrapolate the electronic energy to the complete basis set (CBS) limit. Additional complete-active-space and relativistic corrections were also incorporated into the PES. For the DMS, the MRCI/aug-cc-pwCV5Z level of theory was used, which is known to produce reliable line intensities~\citep{Oleg:2015}.

   To demonstrate that our variational calculations are robust, we also employ a perturbative nonrigid-inverter (NRI) theory approach~\citep{Spirko:1983}, which has previously been used to investigate ammonia~\citep{Spirko:2014,Owens:2015}. The NRI potential energy function for hydronium~\citep{Spirko:1989} was upgraded by fitting to a much broader set of experimental data. 

\section{Results and Discussion}

   The results are illustrated in Figures 1 to 4, with detailed tables provided as supplementary material (extracts of the tables are presented in Tables 1-11). In Fig. 1, the rotational dependence of the sensitivities for the inversion transitions in the ground vibrational state of H$_3{}^{16}$O$^+$ and H$_3{}^{18}$O$^+$ is shown. The non-smooth behaviour of the ($J,K\!=\!3$) transitions is caused by the $\Delta k=\pm 3$ interactions (for details see \citet{Belov:1980}). For D$_3{}^{16}$O$^+$ the sensitivities display a very similar, albeit smoother trend.

\begin{figure*}
\includegraphics[width=0.497\textwidth,angle=0]{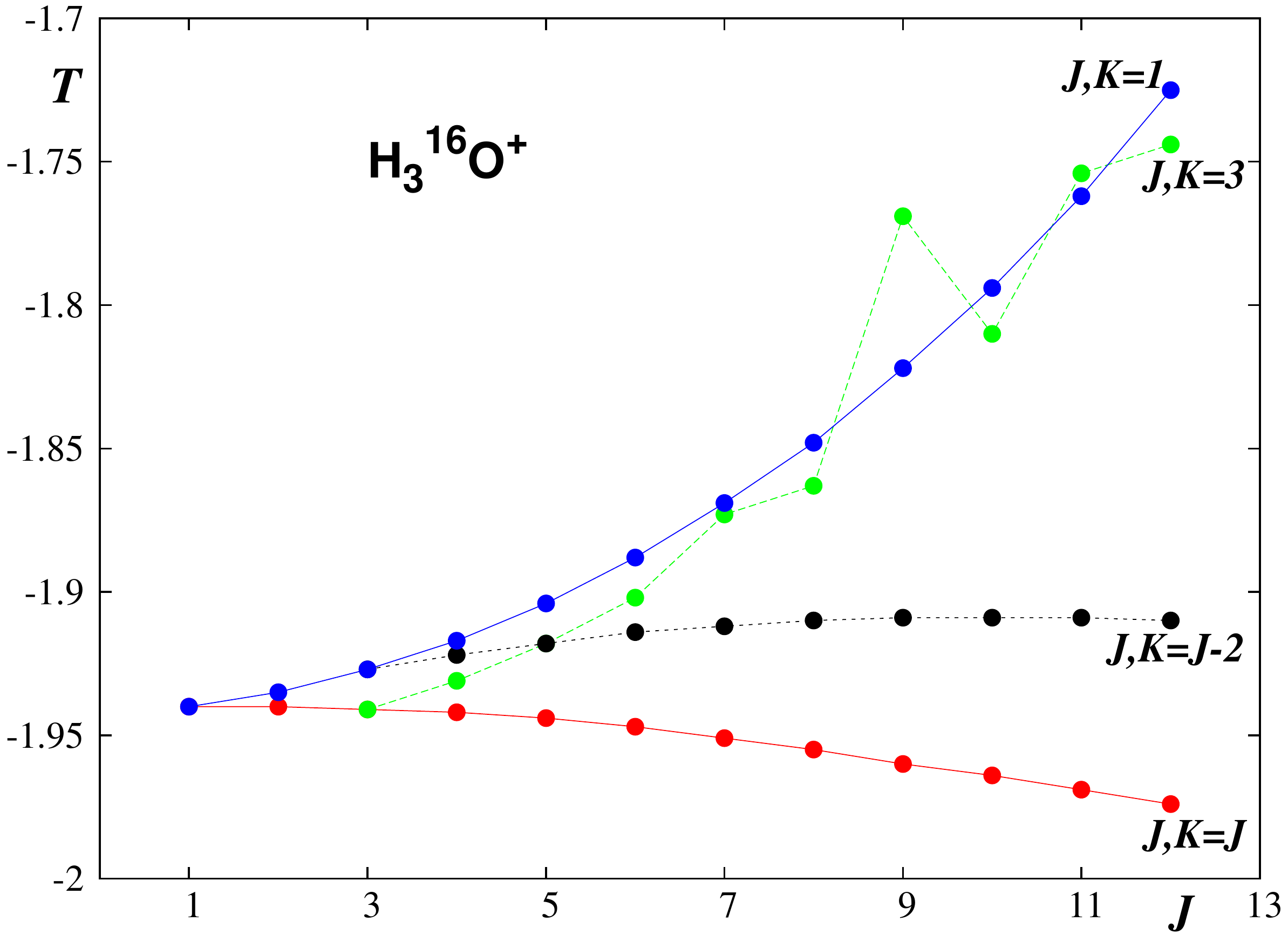}
\includegraphics[width=0.497\textwidth,angle=0]{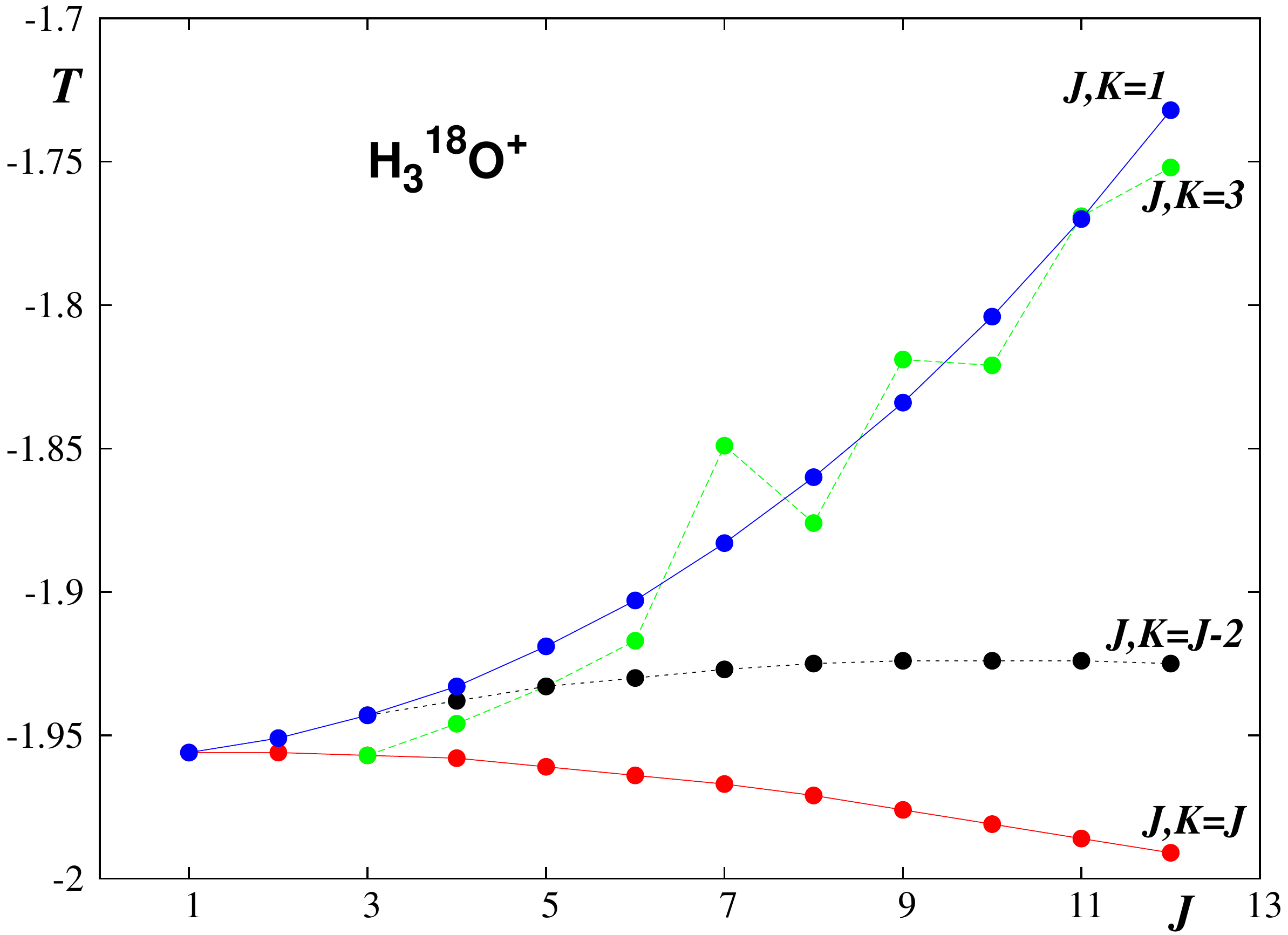}
\caption{Rotational dependence of the sensitivities ($T$) of the inversion transitions in the ground vibrational states of H$_3{}^{16}$O$^+$ and H$_3{}^{18}$O$^+$.}
\end{figure*} 
   
   More encouraging are the low $J$ rotation-inversion transitions displayed in Fig. 2, of which a large number have been observed experimentally in both laboratory~\citep{Yu:2014,Furuya:2005}, and astronomical~\citep{Wootten:1991,Phillips:1992,Goi:2001,Tak:2006,Gonzalez:2013} environments. The appearance of both positive and negative sensitivities is beneficial to constrain a possible variation in the proton-to-electron mass ratio. The effective Hamiltonian model used by \citet{Kozlov:2011b} (KPR in Fig. 2), which does not account for all centrifugal corrections, shows consistent agreement with both the nonrigid-inverter theory (NRI), and variational (TROVE) results. Thus the strongly anomalous sensitivity coefficients of the 1$_{1,1}^-\leftarrow$2$_{2,1}^+$ and 1$_{1,0}^-\leftarrow$2$_{2,0}^+$ transitions of H$_2$DO$^+$, and the 1$_{0,1}^-\leftarrow$1$_{1,1}^+$ transition of HD$_2$O$^+$, proposed by \citet{Kozlov:2011b} have real promise. As discussed previously~\citep{Kozlov:2011b}, the results of \citet{Kozlov:2011a} (KL in Fig. 2) overestimate the H$_3{}^{16}$O$^+$ sensitivities and should not be used in future studies.
   
\begin{figure*}
\includegraphics[width=0.497\textwidth,angle=0]{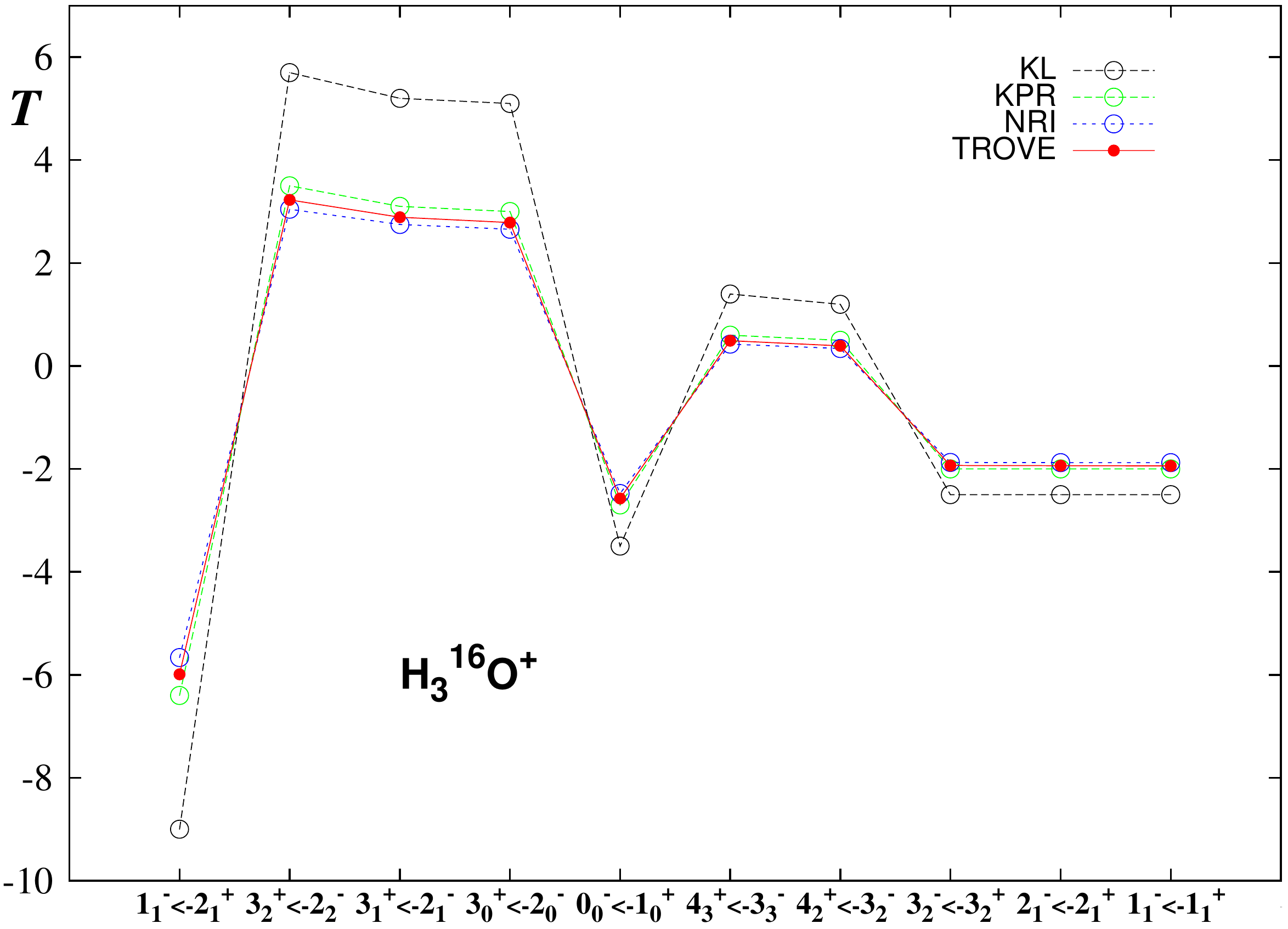}
\includegraphics[width=0.497\textwidth,angle=0]{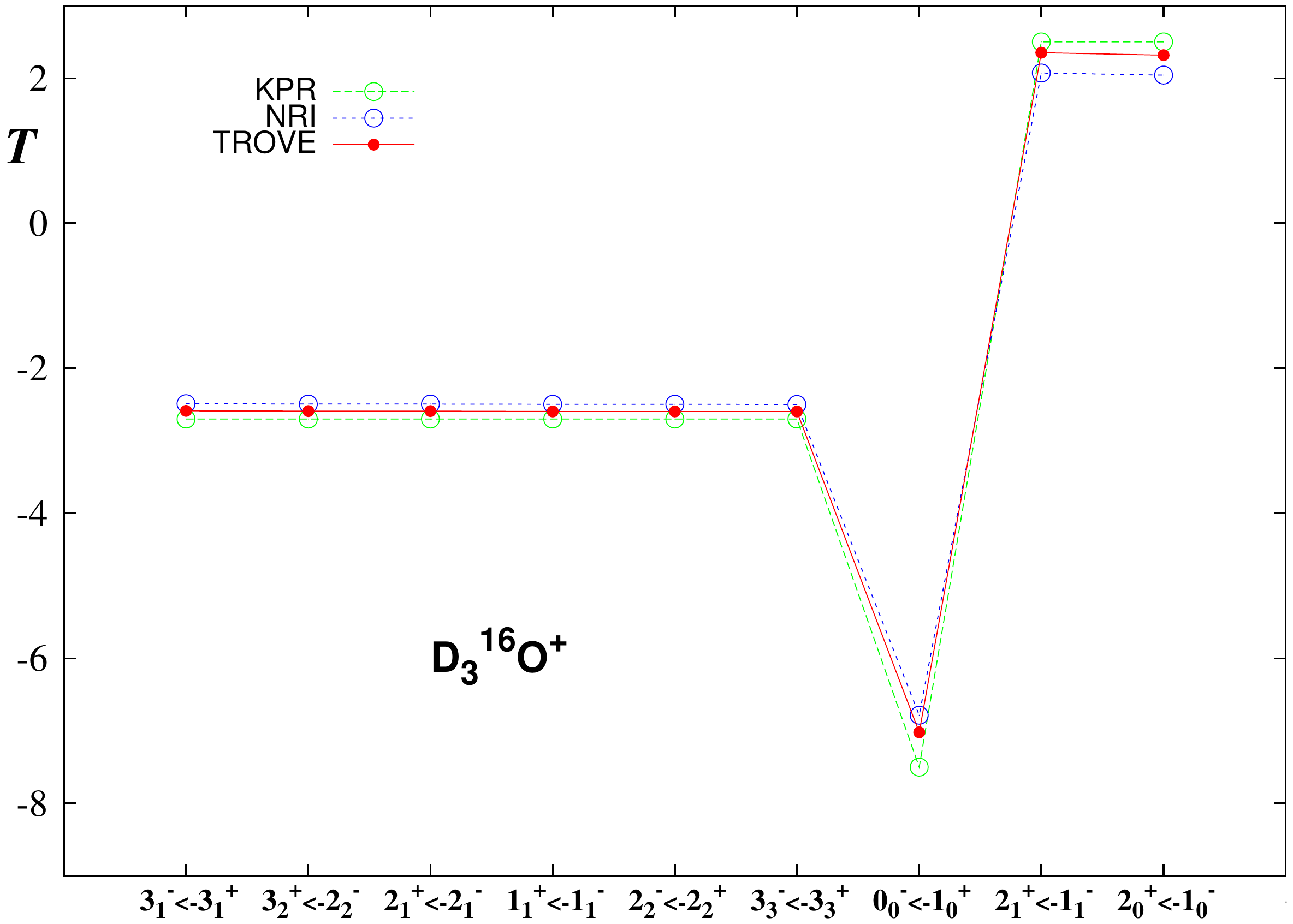}
\caption{State dependence of the calculated sensitivities ($T$) of the rotation-inversion transitions in the ground vibrational states of H$_3{}^{16}$O$^+$ and D$_3{}^{16}$O$^+$. KL: calculated in \citet{Kozlov:2011a}; KPR: calculated in \citet{Kozlov:2011b}; NRI: calculated using the nonrigid-inverter theory (this study); TROVE: calculated variationally (this study). States are labelled as $J_{K}^{\,\pm}$ on the x-axis.}
\end{figure*}   

   The $\Delta k=\pm 3$ interactions give rise to several `forbidden' ro-vibrational combination differences of the $\nu_3$ band  (see Fig. 3). The most sensitive of these are presented in Fig. 4. Notably the 7$_{3}^-\leftarrow$7$_{0}^+$ and 9$_{3}^-\leftarrow$9$_{0}^+$ combination differences, for which a number of the corresponding transitions have been observed experimentally~\citep{Uy:1997}, have theoretically derived values of $T=-15.416$ and $10.518$, respectively. The difference, $\Delta T=25.934$, is comparable to the most stringent limit on $\mu$ obtained using methanol, which utilized transitions with $\Delta T=31.8$~\citep{Bagdonaite:2013}. However, it should be noted that this constraint has recently been deemed unreliable, and subsequently replaced by a more robust value which employed methanol transitions with $\Delta T=6.4$~\citep{Kanekar:2015}. Despite available experimental data~\citep{Araki:1999}, the D$_3{}^{16}$O$^+$ counterparts of these combination differences do not appear to be of any real use, with sensitivities around $T=-1.006$ (see supplementary material for more detail).

\begin{figure}
\centering
\includegraphics[width=1.0\columnwidth]{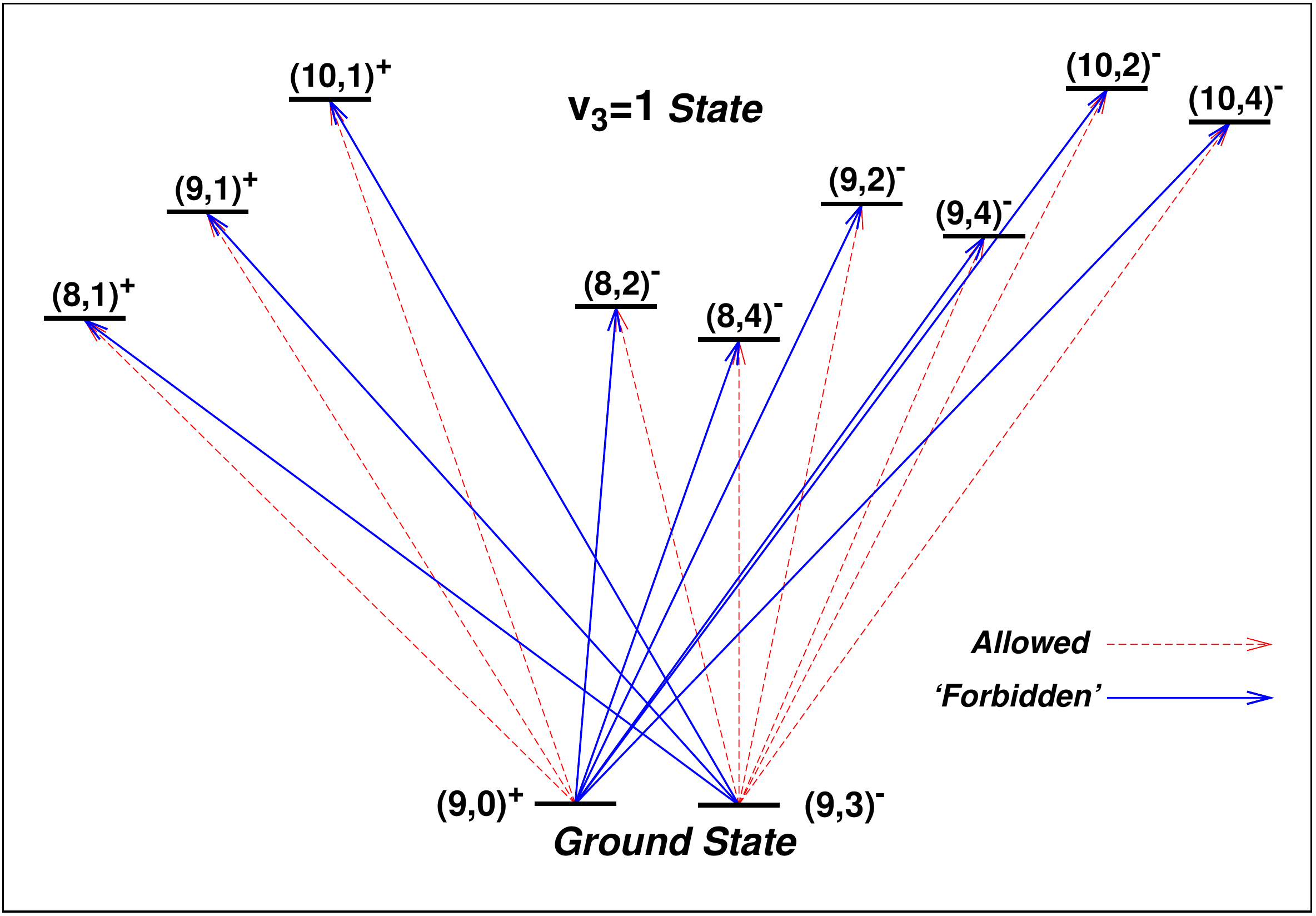}
\caption{ The $|J\!=\!9,K\!=\!0,v_2\!=\!0^+\rangle\!-|J\!=\!9,K\!=\!3,v_2\!=\!0^-\rangle$ combination differences of the $\nu_3$ band of H$_3$O$^+$.}
\end{figure}

\begin{figure}
\centering
\includegraphics[width=1.0\columnwidth]{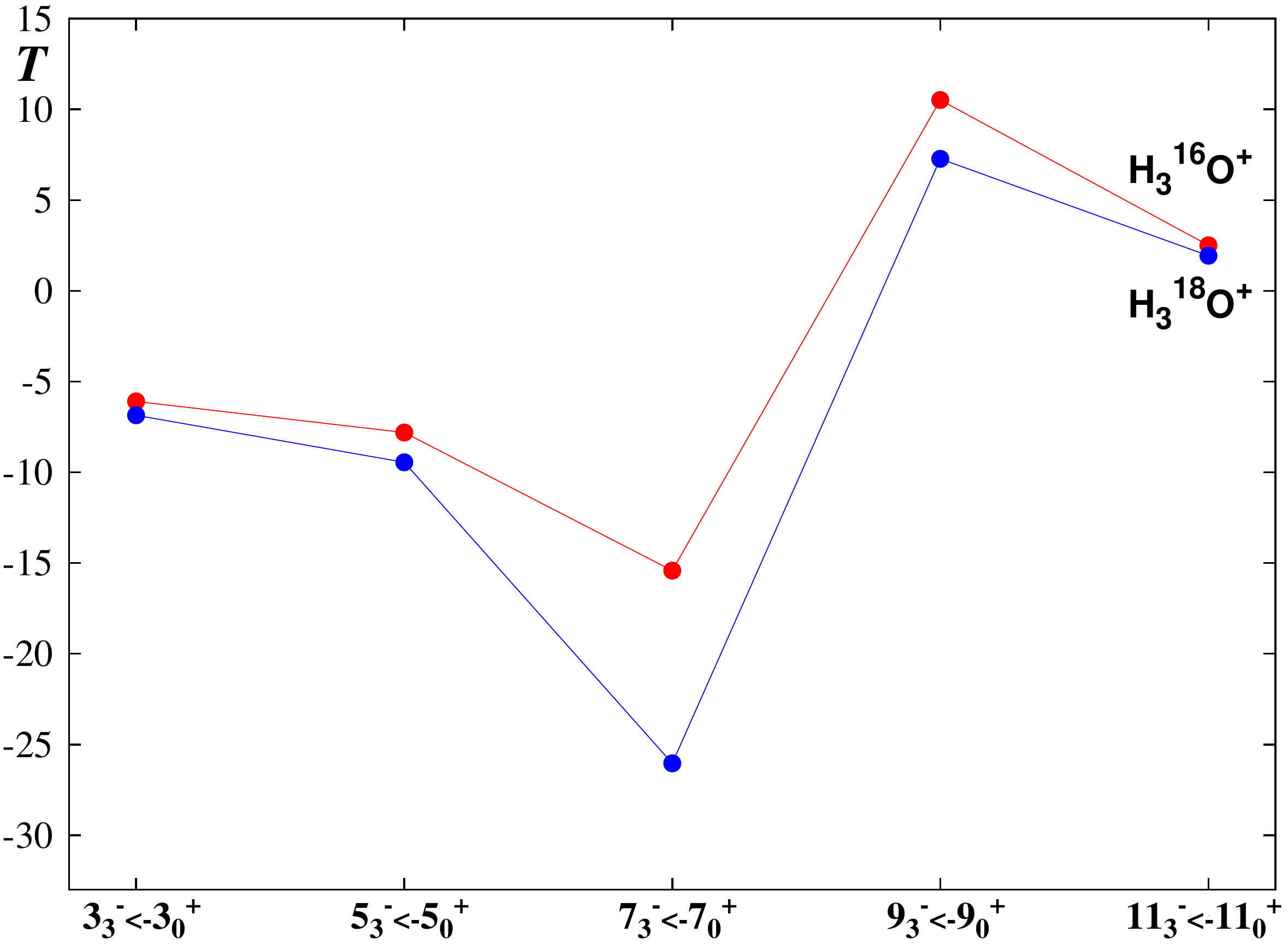}
 \caption{ The sensitivities ($T$) of the most sensitive $|J,K\!=\!0,v_2\!=\!0^+\rangle\!-|J,K\!=\!3,v_2\!=\!0^-\rangle$ $(J=3,5,7,9,11)$ combination differences of H$_3{}^{16}$O$^+$ and H$_3{}^{18}$O$^+$. States are labelled as $J_{K}^{\,\pm}$ on the x-axis.}
\end{figure}

\section{Conclusion}

   A robust variational study of the vibration-rotation-inversion transitions of H$_3{}^{16}$O$^+$, H$_3{}^{18}$O$^+$, and D$_3{}^{16}$O$^+$ has been carried out. We hope that by providing theoretical frequency data and Einstein A coefficients, future laboratory and astronomical observations can be tailored to measure transitions which possess sizeable sensitivities. The astrophysical importance of hydronium suggests that this is a realistic prospect. Emphasis should be placed on the `forbidden' combination differences of the $\nu_3$ band, since several of the corresponding transitions have already been experimentally measured~\citep{Uy:1997}. The 7$_{3}^-\leftarrow$7$_{0}^+$ and 9$_{3}^-\leftarrow$9$_{0}^+$ combination differences are separated by $\Delta T=25.934$. This is around four times larger than the $\Delta T$ of the methanol transitions recently used to determine the most reliable constraint on a possible variation in the proton-to-electron mass ratio~\citep{Kanekar:2015}.
   
\section*{Acknowledgments}
The work was a part of the research project RVO:61388963 (IOCB) and was supported by the Czech Science Foundation (grant P209/15-10267S). AO acknowledges the UCL Impact Studentship scheme. SY thanks ERC Advanced Investigator Project 267219. The State Project No. 0035-2014-009 is acknowledged by RIO.

\bibliographystyle{mn2e}

\section*{Supporting Information}
Additional Supporting Information may be found in the online version of this article:

\begin{table*}
\centering
\vspace*{-0.0cm}
\caption{Inversion frequencies ($\nu$), Einstein coefficients ($A$), and sensitivities ($T$) of H$_3{}^{16}$O$^+$ in the ground vibrational state. The full table is available online as supplementary material.}
\label{tab:inv_ground_h3op}
\begin{tabular}{c 
                r 
                c @{\extracolsep{0.18in}}
                l @{\extracolsep{0.20in}}
                c @{\extracolsep{0.30in}}
                c @{\extracolsep{0.01in}}
                r 
                c @{\extracolsep{0.180in}}
                l @{\extracolsep{0.20in}}
                c}
\hline \\[-2mm]
\multicolumn{1}{c}{$J$}&\multicolumn{1}{c}{$K$}&\multicolumn{1}{c}{$\nu_{\mathrm{calc}}$/GHz}&\multicolumn{1}{c}{$A$/s$^{-1}$}&\multicolumn{1}{c}{$T$}&\multicolumn{1}{c}{$J$}&\multicolumn{1}{c}{$K$}&\multicolumn{1}{c}{$\nu_{\mathrm{calc}}$/GHz}&\multicolumn{1}{c}{$A$/s$^{-1}$}&\multicolumn{1}{c}{$T$}\\[1mm]
\hline \\[-1mm]
   1&  1&  1655.8577& 0.859E-1& -1.940&  9&  4&  1299.1987& 0.156E-1& -1.851\\[-0.3mm]
   2&  1&  1632.1427& 0.275E-1& -1.935&  9&  5&  1361.3791& 0.277E-1& -1.867\\[-0.3mm]
   2&  2&  1657.2795& 0.115E+0& -1.940&  9&  6&  1440.7479& 0.465E-1& -1.887\\[-0.3mm]
   3&  1&  1597.1617& 0.130E-1& -1.927&  9&  7&  1539.6082& 0.758E-1& -1.909\\[-0.3mm]
   3&  2&  1621.8135& 0.540E-1& -1.932&  9&  8&  1660.8177& 0.122E+0& -1.933\\[-0.3mm]
 ...&  ...&  ...& ...& ...& ...& ...&  ...& ...& ...\\[1mm]
\hline \\[-3mm]
\end{tabular}
\end{table*}

\begin{table*}
\centering
\vspace*{-0.0cm}
\caption{Inversion frequencies ($\nu$), Einstein coefficients ($A$), and sensitivities ($T$) of H$_3{}^{18}$O$^+$ in the ground vibrational state. The full table is available online as supplementary material.}
\label{tab:inv_ground_h3o18p}
\begin{tabular}{c 
                r 
                c @{\extracolsep{0.18in}}
                l @{\extracolsep{0.20in}}
                c @{\extracolsep{0.30in}}
                c @{\extracolsep{0.01in}}
                r 
                c @{\extracolsep{0.1800in}}
                l @{\extracolsep{0.20in}}
                c}
\hline \\[-2mm]
\multicolumn{1}{c}{$J$}&\multicolumn{1}{c}{$K$}&\multicolumn{1}{c}{$\nu_{\mathrm{calc}}$/GHz}&\multicolumn{1}{c}{$A$/s$^{-1}$}&\multicolumn{1}{c}{$T$}&\multicolumn{1}{c}{$J$}&\multicolumn{1}{c}{$K$}&\multicolumn{1}{c}{$\nu_{\mathrm{calc}}$/GHz}&\multicolumn{1}{c}{$A$/s$^{-1}$}&\multicolumn{1}{c}{$T$}\\[1mm]
\hline \\[-1mm]
   1&  1&  1608.7744&  0.788E-1& -1.956&  9&  4&  1249.7781& 0.139E-1& -1.863\\[-0.3mm]
   2&  1&  1584.8777&  0.252E-2& -1.951&  9&  5&  1311.5302& 0.248E-1& -1.881\\[-0.3mm]
   2&  2&  1610.0266&  0.105E+0& -1.956&  9&  6&  1390.4774& 0.419E-1& -1.901\\[-0.3mm]
   3&  1&  1549.6465&  0.119E-1& -1.943&  9&  7&  1488.9820& 0.687E-1& -1.924\\[-0.3mm]
   3&  2&  1574.2941&  0.495E-1& -1.948&  9&  8&  1609.9871& 0.111E+0& -1.949\\[-0.3mm]
   ...&  ...&  ...&  ...& ... & ...& ...&  ...& ...& ...\\[1mm]
\hline \\[-3mm]
\end{tabular}
\end{table*}

\begin{table*}
\centering
\vspace*{-0.0cm}
\caption{The rotation-inversion frequencies ($\nu$), Einstein coefficients ($A$), and sensitivities ($T$) of H$_3{}^{16}$O$^+$ in the ground vibrational state$^a$. The full table is available online as supplementary material.}
\begin{tabular}{c
                c @{\extracolsep{0.01in}}
                c @{\extracolsep{0.01in}}
                c @{\extracolsep{0.1000in}}
                c @{\extracolsep{0.01in}}
                c @{\extracolsep{0.01in}}
                c @{\extracolsep{0.0100in}}
                c @{\extracolsep{0.2000in}}
                c @{\extracolsep{0.2000in}}
                c @{\extracolsep{0.1500in}}
                c @{\extracolsep{0.1500in}}
                c @{\extracolsep{0.1500in}}
                c @{\extracolsep{0.0010in}}
                c}
\hline \\[-2mm]
\multicolumn{1}{c}{$\Gamma\p$}&\multicolumn{1}{c}{$p\p$}&\multicolumn{1}{c}{$J\p$}&\multicolumn{1}{c}{$K\p$}&\multicolumn{1}{c}{$\Gamma\pp$}&\multicolumn{1}{c}{$p\pp$}&\multicolumn{1}{c}{$J\pp$}&\multicolumn{1}{c}{$K\pp$}&\multicolumn{1}{c}{$\nu_{\mathrm{calc}}$/MHz}&\multicolumn{1}{c}{$\nu_{\mathrm{exp}}$/MHz}&\multicolumn{1}{c}{$A$/s$^{-1}$}&\multicolumn{1}{c}{$T^b$}&\multicolumn{1}{c}{$T^c$}& \\[1mm]
\hline \\[-1mm]
$E\p$   & 0$^-$&  1&  1& $E\pp$  & 0$^+$&  2&  1&   308483.172&    307192.410$^d$&  0.556E-3&   -5.992&     -6.017\\[0.3mm]
$E\p$   & 0$^+$&  3&  2& $E\pp$  & 0$^-$&  2&  2&   362865.643&    364797.427$^{d,e}$&  0.432E-3&    3.227&      3.210\\[0.3mm]
$E\pp$  & 0$^+$&  3&  1& $E\p$   & 0$^-$&  2&  1&   386507.906&    388458.641&  0.838E-3&    2.891&      2.876\\[0.3mm]
$A_2\p$ & 0$^+$&  3&  0& $A_2\pp$& 0$^-$&  2&  0&   394315.581&    396272.412$^f$&  0.100E-2&    2.788&      2.775\\[0.3mm]
$A_2\pp$& 0$^-$&  0&  0& $A_2\p$ & 0$^+$&  1&  0&   985361.418&    984711.888&  0.362E-1&   -2.575&     -2.577\\[0.3mm]
...  & ...&  ...&  ...& ...   & ...&  ...&  ...&  ...&   ...&  ...&   ...&     ...\\[1mm]
\hline \\[-3mm]
\end{tabular}
\vspace*{1mm}
\\[0mm]
\footnotesize $^a$ If not stated otherwise, the experimental frequencies have been taken from \citet{Yu:2014}. $^b$ Calculated using theoretical frequencies. $^c$ Calculated using experimental frequencies. $^d$ Also observed astronomically in \citet{Tak:2006} $^e$ Also observed astronomically in \citet{Wootten:1991} $^f$ Also observed astronomically in \citet{Phillips:1992} $^g$ Astronomical observation from \citet{Goi:2001}. $^h$ Astronomical observation from \citet{Gonzalez:2013}.\\
\end{table*}

\begin{table*}
\centering
\vspace*{-0.0cm}
\caption{The rotation-inversion frequencies ($\nu$), Einstein coefficients ($A$), and sensitivities ($T$) of H$_3{}^{18}$O$^+$ in the ground vibrational state. The full table is available online as supplementary material.}
\begin{tabular}{c
                c @{\extracolsep{0.01in}}
                c @{\extracolsep{0.01in}}
                c @{\extracolsep{0.1000in}}
                c @{\extracolsep{0.01in}}
                c @{\extracolsep{0.01in}}
                c @{\extracolsep{0.0100in}}
                c @{\extracolsep{0.2000in}}
                c @{\extracolsep{0.1500in}}
                c @{\extracolsep{0.1500in}}
                c}
\hline \\[-2mm]
\multicolumn{1}{c}{$\Gamma\p$}&\multicolumn{1}{c}{$p\p$}&\multicolumn{1}{c}{$J\p$}&\multicolumn{1}{c}{$K\p$}&\multicolumn{1}{c}{$\Gamma\pp$}&\multicolumn{1}{c}{$p\pp$}&\multicolumn{1}{c}{$J\pp$}&\multicolumn{1}{c}{$K\pp$}&\multicolumn{1}{c}{$\nu_{\mathrm{calc}}$/MHz}&\multicolumn{1}{c}{$A$/s$^{-1}$}&\multicolumn{1}{c}{$T$}\\[1mm]
\hline \\[-1mm]
$A_2\pp$& 0$^-$ & 0&  0& $A_2\p$ & 0$^+$  & 1&  0&    939604& 0.314E-1&  -2.633\\[0.3mm]
$E\p$   & 0$^-$ & 2&  1& $E\pp$  & 0$^+$  & 1&  1&   2929768& 0.287E+0&  -1.520\\[0.3mm]
$A_2\pp$& 0$^-$ & 2&  0& $A_2\p$ & 0$^+$  & 1&  0&   2921121& 0.379E+0&  -1.518\\[0.3mm]
$E\p$   & 0$^-$ & 1&  1& $E\pp$  & 0$^+$  & 2&  1&    263884& 0.349E-3&  -6.765\\[0.3mm]
$E\pp$  & 0$^-$ & 3&  2& $E\p$   & 0$^+$  & 2&  2&   3590704& 0.419E+0&  -1.423\\[0.3mm]
... & ... & ...&  ...& ...& ...  & ...&  ...&   ...& ...&  ...\\[1mm]
\hline \\[-3mm]
\end{tabular}
\end{table*}

\begin{table*}
\centering
\vspace*{-0.0cm}
\caption{The frequencies ($\nu$), Einstein coefficients ($A$), and sensitivities ($T$) of the strongest `forbidden' rotation-inversion transitions in the ground vibrational state of H$_3{}^{16}$O$^+$. The full table is available online as supplementary material.}
\label{tab:ro_inv_ground_h3op}
\begin{tabular}{c
                c @{\extracolsep{0.01in}}
                c @{\extracolsep{0.01in}}
                c @{\extracolsep{0.1000in}}
                c @{\extracolsep{0.01in}}
                c @{\extracolsep{0.01in}}
                c @{\extracolsep{0.0100in}}
                c @{\extracolsep{0.2000in}}
                c @{\extracolsep{0.1500in}}
                c @{\extracolsep{0.1500in}}
                c}
\hline \\[-2mm]
\multicolumn{1}{c}{$\Gamma\p$}&\multicolumn{1}{c}{$p\p$}&\multicolumn{1}{c}{$J\p$}&\multicolumn{1}{c}{$K\p$}&
\multicolumn{1}{c}{$\Gamma\pp$}&\multicolumn{1}{c}{$p\pp$}&\multicolumn{1}{c}{$J\pp$}&\multicolumn{1}{c}{$K\pp$}&\multicolumn{1}{c}{$\nu_{\mathrm{calc}}$/MHz}&
\multicolumn{1}{c}{$A$/s$^{-1}$}&\multicolumn{1}{c}{$T$}\\[1mm]
\hline \\[-1mm]
$A_2\p$ & 0$^+$&  7& 0& $A_2\pp$& 0$^+$&  6& 3&  6013041& 0.532E-1& -1.012\\[1mm]
$A_2\p$ & 0$^-$&  7& 3& $A_2\pp$& 0$^-$&  6& 0&  3320905& 0.115E-1& -0.982\\[1mm]
$A_2\pp$& 0$^+$&  8& 3& $A_2\p$ & 0$^+$&  7& 0&  3986710& 0.132E-1& -0.985\\[1mm]
$A_2\pp$& 0$^-$&  8& 0& $A_2\p$ & 0$^-$&  7& 3&  6509745& 0.713E-1& -0.980\\[1mm]
$A_2\p$ & 0$^+$&  9& 0& $A_2\pp$& 0$^+$&  8& 3&  7288334& 0.473E+0& -1.011\\[1mm]
...   & ...&  ...& ...& ...  & ...&  ...& ...&  ...& ...& ...\\[1mm]
\hline\\[-3mm]
\end{tabular}
\end{table*}

\begin{table*}
\centering
\vspace*{-0.0cm}
\caption{Combination differences ($CD$) of the `forbidden' ($\Delta|k-l|=3$) and allowed ($\Delta|k-l|=0$) transitions between the $\nu_3$ and ground vibrational states in H$_3{}^{16}$O$^+${\,}$^a$\,. The full table is available online as supplementary material.}
\label{tab:cd_h3op}
\begin{tabular}{c
                c @{\extracolsep{0.001in}}
                c @{\extracolsep{0.001in}}
                c @{\extracolsep{0.150in}}
                c @{\extracolsep{0.001in}}
                c @{\extracolsep{0.001in}}
                c @{\extracolsep{0.005in}}
                c @{\extracolsep{0.150in}}
                c}
\hline \\[-2mm]
\multicolumn{1}{c}{Allowed}&\multicolumn{1}{c}{$\nu_{\mathrm{calc}}$/cm$^{-1}$}&\multicolumn{1}{c}{$\nu_{\mathrm{exp}}$/cm$^{-1}$}&\multicolumn{1}{c}{$A$/s$^{-1}$}&\multicolumn{1}{c}{Forbidden}&\multicolumn{1}{c}{$\nu_{\mathrm{calc}}$/cm$^{-1}$}&\multicolumn{1}{c}{$\nu_{\mathrm{exp}}$/cm$^{-1}$}&\multicolumn{1}{c}{$A$/s$^{-1}$}&\multicolumn{1}{c}{$CD$/cm$^{-1}$} \\[1mm]
\hline \\[-1mm]
 $^r$P(3,0)$^+$ & 3457.025&          & 0.390E+3& $^o$P(3,3)$^-$ & 3447.266&          & 0.826E-1& 9.7594\\[0.3mm]
 $^r$Q(3,0)$^+$ & 3523.544&          & 0.964E+3& $^o$Q(3,3)$^-$ & 3513.785&          & 0.189E-1& 9.7594\\[0.3mm]
 $^r$R(3,0)$^+$ & 3610.441&          & 0.530E+3& $^o$R(3,3)$^-$ & 3600.682&          & 0.326E-1& 9.7594\\[0.3mm]
 $^p$P(3,3)$^-$ & 3474.787&          & 0.934E+3& $^s$P(3,0)$^+$ & 3484.546&          & 0.420E-1& 9.7594\\[0.3mm]
 $^p$Q(3,3)$^-$ & 3539.922&          & 0.233E+3& $^s$Q(3,0)$^+$ & 3549.681&          & 0.142E+0& 9.7594\\[0.3mm]
 $^p$R(3,3)$^-$ & 3626.725&          & 0.246E+2& $^s$R(3,0)$^+$ & 3636.484&          & 0.222E+0& 9.7594\\[0.3mm]
 $^r$R(3,3)$^-$ & 3564.692&          & 0.730E+3& $^u$R(3,0)$^+$ & 3574.452&          & 0.104E-1& 9.7594\\[0.3mm]
 ... & ...&    ...      & ...& ... & ...&    ...      & ...& ...\\[1mm]
\hline \\[-3mm]
\end{tabular}
\vspace*{1mm}
\\[0mm]
\footnotesize $^a$ Experimental frequencies from \citet{Tang:1999} and \citet{Uy:1997}. Experimental $CD$ data in parentheses. Transitions with $\Delta J=-1,0,+1$ are described using the labels P, Q, R respectively, whilst the superscript $o,p,q,r,s,t,u$ notation corresponds to transitions with $\Delta K=-2,-1,0,+1,+2,+3,+4$ respectively. All transitions are between states of $A_2\p$ and $A_2\pp$ symmetry, where $+(-)\rightarrow+(-)$ are allowed, and $+(-)\rightarrow-(+)$ are forbidden. See also Figure 3. \\
\end{table*}

\begin{table*}
\centering
\vspace*{-0.0cm}
\caption{The `forbidden' combination differences ($\nu$) and sensitivities ($T$) of the H$_3{}^{16}$O$^+$ and H$_3{}^{18}$O$^+$ ground vibrational state transitions$^a$\,. The full table is available online as supplementary material.}
\begin{tabular}{c
                c @{\extracolsep{0.002in}}
                c @{\extracolsep{0.002in}}
                c @{\extracolsep{0.180in}}
                c @{\extracolsep{0.002in}}
                c @{\extracolsep{0.002in}}
                c @{\extracolsep{0.002in}}
                c @{\extracolsep{0.09800in}}
                c @{\extracolsep{0.09800in}}
                c @{\extracolsep{0.09800in}}
                c @{\extracolsep{0.09800in}}
                c @{\extracolsep{0.09800in}}
                c}
\hline \\[-2mm]
\multicolumn{1}{c}{$\Gamma\p$}&\multicolumn{1}{c}{$p\p$}&\multicolumn{1}{c}{$J\p$}&\multicolumn{1}{c}{$K\p$}&\multicolumn{1}{c}{$\Gamma\pp$}&\multicolumn{1}{c}{$p\pp$}&\multicolumn{1}{c}{$J\pp$}&\multicolumn{1}{c}{$K\pp$}&\multicolumn{1}{c}{$\nu_{\mathrm{calc}}$/MHz}&\multicolumn{1}{c}{$\nu_{\mathrm{exp}}$/MHz}&\multicolumn{1}{c}{$T^b$}&\multicolumn{1}{c}{$T^c$}& \\[1mm]
\hline \\[-1mm]
 & & & & & & & & H$_3{}^{16}$O$^+$ & & &  \\[1mm]
$A_2\pp$& 0$^+$&  8&  3& $A_2\pp$& 0$^-$&  8&  6&  2490592& 2499819&  -0.492&  -0.490\\[0.3mm]
$A_2\pp$& 0$^+$&  9&  3& $A_2\pp$& 0$^-$&  9&  6&  2549767& 2557200&  -0.536&  -0.534\\[0.3mm]
$E\p$   & 0$^+$&  7&  4& $E\p$   & 0$^-$&  7&  7&  3257694& 3261952&  -0.566&  -0.565\\[0.3mm]
$E\p$   & 0$^+$&  8&  4& $E\p$   & 0$^-$&  8&  7&  3311613& 3316064&  -0.597&  -0.596\\[0.3mm]
$E\p$   & 0$^+$&  8&  4& $E\p$   & 0$^-$&  8&  7&  3311613& 3316124&  -0.597&  -0.596\\[0.3mm]
... & ...& ...&  ...& ... & ...& ...&  ...&   ...&   ...     &   ...& ...\\[1mm]
\hline \\[-3mm]
\end{tabular}
\vspace*{1mm}
\\[0mm]
\footnotesize $^a$ Experimental frequencies from \citet{Tang:1999} and \citet{Uy:1997}. $^b$ Calculated using theoretical frequencies. $^c$ Calculated using experimental frequencies.
\end{table*}

\begin{table*}
\centering
\vspace*{-0.0cm}
\caption{Inversion frequencies ($\nu$), Einstein coefficients ($A$), and sensitivities ($T$) of D$_3{}^{16}$O$^+$ in the ground vibrational state. The full table is available online as supplementary material.}
\label{tab:inv_ground_d3op}
\begin{tabular}{c 
                r 
                c @{\extracolsep{0.08in}}
                c @{\extracolsep{0.20in}}
                c @{\extracolsep{0.30in}}
                c @{\extracolsep{0.01in}}
                r 
                c @{\extracolsep{0.0800in}}
                c @{\extracolsep{0.20in}}
                c}
\hline \\[-2mm]
\multicolumn{1}{c}{$J$}&\multicolumn{1}{c}{$K$}&\multicolumn{1}{c}{$\nu_{\mathrm{calc}}$/MHz}&\multicolumn{1}{c}{$A$/s$^{-1}$}&\multicolumn{1}{c}{$T$}&\multicolumn{1}{c}{$J$}&\multicolumn{1}{c}{$K$}&\multicolumn{1}{c}{$\nu_{\mathrm{calc}}$/MHz}&\multicolumn{1}{c}{$A$/s$^{-1}$}&\multicolumn{1}{c}{$T$}\\[1mm]
\hline \\[-1mm]
   1&  1&    461457.7&  0.202E-2& -2.594&   9&  3&    396223.5&  0.262E-3& -2.532\\[-0.3mm]
   2&  1&    457746.8&  0.659E-3& -2.591&   9& -3&    396307.6&  0.262E-3& -2.533\\[-0.3mm]  
   2&  2&    462036.6&  0.271E-2& -2.595&   9&  4&    404995.9&  0.495E-3& -2.541\\[-0.3mm]
   3&  1&    452238.3&  0.318E-3& -2.586&   9&  5&    416478.0&  0.837E-3& -2.552\\[-0.3mm]
   3&  2&    456477.0&  0.131E-2& -2.590&   9&  6&    430926.5&  0.133E-2& -2.565\\[-0.3mm]
   ...&  ...&    ...&  ...& ...&  ...& ...&    ...&  ...& ...\\[1mm] 
\hline \\[-3mm]
\end{tabular}
\end{table*}

\begin{table*}
\centering
\vspace*{-0.0cm}
\caption{The frequencies ($\nu$), Einstein coefficients ($A$), and sensitivities ($T$) of the rotation-inversion transitions in the ground vibrational state of D$_3{}^{16}$O$^+$. The full table is available online as supplementary material.}
\label{tab:ro_inv_ground_d3op}
\begin{tabular}{c
                c @{\extracolsep{0.01in}}
                c @{\extracolsep{0.01in}}
                c @{\extracolsep{0.1000in}}
                c @{\extracolsep{0.01in}}
                c @{\extracolsep{0.01in}}
                c @{\extracolsep{0.0100in}}
                c @{\extracolsep{0.2000in}}
                c @{\extracolsep{0.1500in}}
                c @{\extracolsep{0.1500in}}
                c}
\hline \\[-2mm]
\multicolumn{1}{c}{$\Gamma\p$}&\multicolumn{1}{c}{$p\p$}&\multicolumn{1}{c}{$J\p$}&\multicolumn{1}{c}{$K\p$}&
\multicolumn{1}{c}{$\Gamma\pp$}&\multicolumn{1}{c}{$p\pp$}&\multicolumn{1}{c}{$J\pp$}&\multicolumn{1}{c}{$K\pp$}&\multicolumn{1}{c}{$\nu_{\mathrm{calc}}$/MHz}&\multicolumn{1}{c}{$A$/s$^{-1}$}&\multicolumn{1}{c}{$T$} \\[1mm]
\hline \\[-1mm]
$A_1\pp$& 0$^-$& 1&  0& $A_1\p$ & 0$^+$& 0&  0&      799894$^a$&     0.703E-2&  -1.919\\[0.3mm]
$A_2\pp$& 0$^-$& 0&  0& $A_2\p$ & 0$^+$& 1&  0&      122016&     0.748E-4&  -7.018\\[0.3mm]
$E\p$   & 0$^-$& 2&  1& $E\pp$  & 0$^+$& 1&  1&     1137348&     0.182E-1&  -1.644\\[0.3mm]
$A_2\pp$& 0$^-$& 2&  0& $A_2\p$ & 0$^+$& 1&  0&     1135859&     0.242E-1&  -1.643\\[0.3mm]
$E\pp$  & 0$^+$& 2&  1& $E\p$   & 0$^-$& 1&  1&      218144&     0.128E-3&   2.352\\[0.3mm]
... & ...& ...&  ...& ...& ...& ...&  ...&     ...&     ...&  ...\\[1mm]
\hline \\[-3mm]
\end{tabular}
\vspace*{1mm}
\\[0mm]
\footnotesize $^a$ Experimental value of $798713.814{\,}$MHz measured in \citet{Furuya:2005}. Note that states with $K=+3$ are of $A_2$ symmetry, whilst those with $K=-3$ are of $A_1$ symmetry.\\
\end{table*}

\begin{table*}
\centering
\vspace*{-0.0cm}
\caption{The frequencies ($\nu$), Einstein coefficients ($A$), and sensitivities ($T$) of the strongest `forbidden' rotation-inversion transitions in the ground vibrational state of D$_3{}^{16}$O$^+$. The full table is available online as supplementary material.}
\label{tab:forb_ro_inv_ground_d3op}
\begin{tabular}{c
                c @{\extracolsep{0.01in}}
                c @{\extracolsep{0.01in}}
                c @{\extracolsep{0.1000in}}
                c @{\extracolsep{0.01in}}
                c @{\extracolsep{0.01in}}
                c @{\extracolsep{0.0100in}}
                c @{\extracolsep{0.2000in}}
                c @{\extracolsep{0.1500in}}
                c @{\extracolsep{0.1500in}}
                c}
\hline \\[-2mm]
\multicolumn{1}{c}{$\Gamma\p$}&\multicolumn{1}{c}{$p\p$}&\multicolumn{1}{c}{$J\p$}&\multicolumn{1}{c}{$K\p$}&
\multicolumn{1}{c}{$\Gamma\pp$}&\multicolumn{1}{c}{$p\pp$}&\multicolumn{1}{c}{$J\pp$}&\multicolumn{1}{c}{$K\pp$}&\multicolumn{1}{c}{$\nu_{\mathrm{calc}}$/MHz}&\multicolumn{1}{c}{$A$/s$^{-1}$}&\multicolumn{1}{c}{$T$} \\[1mm]
\hline \\[-1mm]
$A_1\pp$& 0$^-$&   9&  0& $A_1\p$ & 0$^-$&   8& -3&  3688528& 0.170E-3& -0.989\\[0.3mm]
$A_1\p$ & 0$^+$&  10&  0& $A_1\pp$& 0$^+$&   9& -3&  4042517& 0.147E-3& -1.000\\[0.3mm]
$A_2\pp$& 0$^-$&  10&  0& $A_2\p$ & 0$^-$&   9&  3&  4016397& 0.375E-3& -0.987\\[0.3mm]
$E\p$   & 0$^-$&  11&  1& $E\pp$  & 0$^-$&  10&  4&  4785669& 0.151E-3& -0.985\\[0.3mm]
$A_2\p$ & 0$^+$&  11&  0& $A_2\pp$& 0$^+$&  10&  3&  4368847& 0.288E-3& -0.997\\[0.3mm]
... & ...&  ...& ...& ...& ...&  ...&  ...&  ...& ...& ...\\[1mm]
\hline \\[-3mm]
\end{tabular}
\vspace*{1mm}
\\[0mm]
\footnotesize Note that states with $K=+3$ are of $A_2$ symmetry, whilst those with $K=-3$ are of $A_1$ symmetry.\\
\end{table*}

\begin{table*}
\centering
\vspace*{-0.0cm}
\caption{The `forbidden' combination differences ($\nu$) and sensitivities ($T$) of the D$_3{}^{16}$O$^+$ ground vibrational state transitions$^a$\,. The full table is available online as supplementary material.}
\label{tab:forb_cd_ground_d3op}
\begin{tabular}{c
                c @{\extracolsep{0.01in}}
                c @{\extracolsep{0.01in}}
                c @{\extracolsep{0.1000in}}
                c @{\extracolsep{0.01in}}
                c @{\extracolsep{0.01in}}
                c @{\extracolsep{0.0100in}}
                c @{\extracolsep{0.2000in}}
                c @{\extracolsep{0.2000in}}
                c @{\extracolsep{0.1500in}}
                c @{\extracolsep{0.1500in}}
                c}
\hline \\[-2mm]
\multicolumn{1}{c}{$\Gamma\p$}&\multicolumn{1}{c}{$p\p$}&\multicolumn{1}{c}{$J\p$}&\multicolumn{1}{c}{$K\p$}&
\multicolumn{1}{c}{$\Gamma\pp$}&\multicolumn{1}{c}{$p\pp$}&\multicolumn{1}{c}{$J\pp$}&\multicolumn{1}{c}{$K\pp$}&
\multicolumn{1}{c}{$\nu_{\mathrm{calc}}$/MHz}&\multicolumn{1}{c}{$\nu_{\mathrm{exp}}$/MHz}&\multicolumn{1}{c}{$T^b$}&\multicolumn{1}{c}{$T^c$} \\[1mm]
\hline \\[-1mm]
$A_2\p$ & 0+&   8&  6& $A_2\p$ & 0+&   7&  6&   2711462&  2714369&  -1.004& -1.003 \\[0.3mm]
$A_2\p$ & 0+&   7&  6& $A_2\p$ & 0+&   6&  6&   2376103&  2378622&  -1.006& -1.005 \\[0.3mm]
$E\p$   & 0+&   6&  4& $E\p$   & 0+&   5&  4&   2035287&  2037351&  -1.005& -1.004 \\[0.3mm]
$A_2\pp$& 0+&   4&  3& $A_2\pp$& 0+&   3&  3&   1358641&  1360071&  -1.006& -1.005 \\[0.3mm]
$A_1\pp$& 0+&   4& -3& $A_1\pp$& 0+&   3& -3&   1358642&  1360071&  -1.006& -1.005 \\[0.3mm]
...   & ...&   ...&  ...& ...   & ...&   ...&  ...&   ...&  ...&  ...& ... \\[1.0mm]
\hline \\[-3mm]
\end{tabular}
\vspace*{1mm}
\\[0mm]
\footnotesize $^a$ Experimental frequencies from \citet{Araki:1999}. $^b$ Calculated using theoretical frequencies. $^c$ Calculated using experimental frequencies.\\
\end{table*}

\label{lastpage}

\end{document}